\documentclass[aps,nofootinbib,prd,eqsecnum,twocolumn,showpacs,showkeys,preprintnumbers]{revtex4-2}
\usepackage[caption=false]{subfig}
\usepackage{graphicx}
\usepackage{amsmath}
\usepackage{amsfonts}
\usepackage{amssymb}
\usepackage{color}
\usepackage{bm}
\usepackage{mathrsfs}
\usepackage{epstopdf}
\usepackage{url}
\usepackage{footnote}
\usepackage{textcomp}
\usepackage{ulem}
\usepackage{esint}
\usepackage{calrsfs}
\DeclareMathAlphabet{\pazocal}{OMS}{zplm}{m}{n}

\newcommand{\Lb}{\pazocal{L}}
\usepackage[unicode=true, pdfusetitle,
bookmarks=true,bookmarksnumbered=false,bookmarksopen=false,
 breaklinks=false,pdfborder={0 0 1},backref=false,colorlinks=false]{hyperref}
\usepackage{multirow}
\usepackage{pifont}
\usepackage{booktabs}
\begin{document}
\title{Observational constraints on the growth index parameters in $f(Q)$ gravity}
%\author{Dalale Mhamdi\textsuperscript{1}}
%\email{dalale.mhamdi@ump.ac.ma}
%\author{Safae Dahmani\textsuperscript{1}}
%\email{dahmani.safae.1026@gmail.com}
%\author{Amine Bouali\textsuperscript{1,2}}
%\email{a1.bouali@ump.ac.ma}
%\author{Imad El Bojaddaini\textsuperscript{1}}
%\email{i.elbojaddaini@ump.ac.ma}
%\author{Taoufik Ouali\textsuperscript{1}}
%\email{t.ouali@ump.ac.ma}
%\date{\today}
%\affiliation{\textsuperscript{1}Laboratory of Physics of Matter and Radiations, Mohammed I University, BP 717, Oujda, Morocco}
%\affiliation{\textsuperscript{2}Higher School of Education and Training, Mohammed I University, BP 717, Oujda, Morocco}
\author{Dalale Mhamdi\textsuperscript{1}}
\email{dalale.mhamdi@ump.ac.ma}
\author{Safae Dahmani\textsuperscript{1}}
\email{dahmani.safae.1026@gmail.com}
\author{Amine Bouali\textsuperscript{1,2}}
\email{a1.bouali@ump.ac.ma}
\author{Imad El Bojaddaini\textsuperscript{1}}
\email{i.elbojaddaini@ump.ac.ma}
\author{Taoufik Ouali\textsuperscript{1}}
\email{t.ouali@ump.ac.ma}
\date{\today}
\affiliation{\textsuperscript{1}Laboratory of Physics of Matter and Radiations, Mohammed I University, BP 717, Oujda, Morocco,\\  \textsuperscript{2}Higher School of Education and Training, Mohammed I University, BP 717, Oujda, Morocco}
	\begin{abstract}

In this study, we analyse constraints on the growth index of matter perturbations, $\gamma$, within the framework of $f(Q)$ gravity, using recent cosmological observations, at the background and the perturbation levels, including Pantheon$^{+}$, Cosmic Chronometer (CC), and Redshift Space Distortion (RSD) datasets. Our analysis focuses on quantifying the distortion parameter, which measures the deviation of the $f(Q)$ gravity model from the concordance $\Lambda$CDM cosmology at the background level. Specifically, we investigate two cases of the growth index parameter: a constant $\gamma$ and a time-varying $\gamma(z)$. We investigate various parametrizations of the growth index $\gamma$, expressed as $\gamma = \gamma_{0} +\gamma_{1} y(z)$, where the function $y(z)$ assumes different forms, including constant ($\Gamma_{0}$), Taylor expansion around $z = 0$ ($\Gamma_{1}$), Taylor expansion around the scale factor ($\Gamma_{2}$), and an exponential form ($\Gamma_{3}$). By employing the Akaike Information Criterion and Bayesian Information Criterion, we find that the combined Pantheon$^{+}$+ CC+ RSD datasets impose stringent constraints on the value of the growth index. For the $\Gamma_{0}$ model, our results indicate that within the concordance $\Lambda$CDM model, $\gamma$ is constrained to $0.545 \pm 0.096$, showing strong agreement with the theoretical expectation of $\gamma_{\Lambda} = \frac{6}{11}$. However, within the framework of $f(Q)$ gravity, we observe $\gamma = 0.571^{+0.095}_{-0.110}$, slightly exceeding the $\Lambda$CDM value by 4.66 $\%$. Furthermore, when considering a time-varying growth index, our analysis reveals that the range of $\gamma_{0}$ spans from $0.596$ to $0.62$ across the $\Gamma_{1-3}$ models. Finally, we compare the obtained results for the growth index parameters within $f(Q)$ with those from previous studies on modified gravity theories, such as $f(R)$ and $f(T)$.

\end{abstract}
	\maketitle
          \tableofcontents
\newpage

\section{Introduction} \label{sec:3}

The prevailing cosmological paradigm, described by the cosmological constant, $\Lambda$, and the Cold Dark Matter (abbreviated as $\Lambda$CDM), has provided a remarkably successful framework for understanding a wide range of cosmological phenomena. From the intricate patterns imprinted on the Cosmic Microwave Background (CMB) radiation \cite{4,5} to the vast web of galaxies and galaxy clusters, $\Lambda$CDM accurately describes the large-scale structure formation \cite{6,7}. Furthermore, $\Lambda$CDM incorporates seamlessly the concept of cosmic expansion \cite{11,22,T3}, elucidating the accelerated rate driven by dark energy, denoted by $\Lambda$.\\

However, despite its successes, the $\Lambda$CDM model is not without its challenges. Chief among these are coincidence \cite{coincidence, coincidence2} and hierarchical \cite{hierarchical,hierarchical1} issues. Additionally, cosmological tension such as the Hubble tension may question the credibility of $\Lambda$CDM as a standard model. This persistent discrepancy between the values of the Hubble constant derived from early and late Universe observations has become a focal point of debate within the cosmological community, for more detail on this issue, we refer the reader to \cite{valentino2021}. Previous works of the authors of this paper have delved into this tension problem, such as the study of $H_0$ tension within a dynamical dark energy model \cite{dahmani2}, and the investigations of the influence of neutrino properties \cite{dahmani1}.
To address the phenomenon of accelerated expansion, scientists have introduced an exotic form of energy density within the framework of general relativity, dubbed dark energy (DE). This has prompted the development of various models, including quintessence \cite{martin2008,ladghami2024}, phantom \cite{Bouali1,Mhamdi1}, and holography \cite{belkacemi2020interacting,bargach2021dynamical,bouhmadi2018more}. These approaches present intriguing avenues for reconciling observational data with theoretical predictions, offering new perspectives on the underlying dynamics of the Universe’s expansion and structure formation.\\

Another approach to explain the observed acceleration of the late-time Universe is theories of modified gravity. Among these Brans-Dicke \cite{1,2,3} modified gravity theories, $f(R)$ gravity introduces modifications to Einstein’s general theory of relativity by incorporating a Lagrangien dependent on the scalar curvature $R$ \cite{De}. This framework has demonstrated success in accounting for the Universe’s accelerating expansion without invoking the concept of dark energy. Additionally, the Teleparallel Equivalent to General Relativity (TEGR) theory, extended to $f(T)$ gravity, proposes an alternative description of gravity based on torsion instead of curvature \cite{Buchdahl}. However, challenges such as violations of local Lorentz invariance have been raised in TEGR gravity. Another intriguing avenue is Symmetric Teleparallel Gravity extended to $f(Q)$ which uses non-metricity $Q$ \cite{JB}. The $f(Q)$ modified theory of gravity has recently been extensively studied in several scenarios \cite{JB,ref12,ref13,ref14} (see also the recent Review on $f(Q)$  gravity in \cite{Heisenberg}). In addition, it was successfully confronted with various observational data, such as the supernovae type Ia, CMB, baryonic acoustic oscillations, growth data, and Redshift Space Distortion \cite{ref21,ref22,ref23,ref24,Koussour}.\\

In this context, understanding the growth of cosmic structures, such as galaxies and galaxy clusters, takes on heightened significance. The growth rate of these structures, quantified by the growth index, serves as a sensitive probe of cosmological models and theories of gravity \cite{Ayuso, Khyllep}. This article specifically aims to investigate the growth of cosmic structures within the framework of $f(Q)$ gravity, characterized by the form $f(Q)=\alpha+\beta Q^n$, as reconstructed by Capozziello and D’Agostino \cite{Capozziello}, through Bayesian analysis employing various cosmological probes at both background and perturbation levels. For instance, in our knowledge, the growth index parameter has been constrained for modified gravity models such as $f(T)$ in \cite{ft} and $f(R)$ gravity in \cite{fr}. However, this analysis has not yet been applied to $f(Q)$ gravity. Thus, the primary aim of this study is to constrain the growth index parameter within the framework of $f(Q)$ gravity and ascertain its expression in two cases: a constant value, and a variable form expressed as $\gamma(z)=\gamma_{0}+\gamma_{1} y(z)$ with the following form of $y(z) = z, y(z) = z/(1 + z),$ and $y(z) =ze^{-z}.$  we present a comprehensive analysis of the growth index and its implications for the viability of $f(Q)$ gravity as a cosmological model. Leveraging both theoretical calculations and observational data, we aim to shed light on the nature of gravity and its role in shaping the large-scale structure of the Universe. Furthermore, The growth index parameter provides a means to distinguish between General Relativity and modified gravity theories, prompting numerous comparative studies \cite{fr,ft}. In our previous paper \cite{Mhamdi2}, we have explored the same $f(Q)$ model, where the growth index parameter was constrained numerically by solving the overdensity equation. In contrast, this study takes a novel approach by considering $\gamma$ as a free parameter, where we use $f=\Omega_{m}^{\gamma}$ parameterization to determine the constraints on the growth index parameters $\gamma_{0}$ and $\gamma_{1}$. The next step is to compare our results with the $\Lambda$CDM model using statistical criteria like the Akaike Information Criterion (AIC) \cite{AIC,akaike1974new} and the Bayesian Information Criterion (BIC) \cite{BIC} to  provides insights into the viability of $f(Q)$ gravity as a cosmological model.\\

The remainder of this paper is organized as follows: In section \ref{$f(Q)$ framework}, we provide an overview of $f(Q)$ gravity and its theoretical foundations. Section \ref{matter perturbations} introduces the linear matter perturbations. The observational data used for both background and perturbation are detailed in section \ref{statistics}. In section \ref{Resullts}, we present our results and discuss their implications for $f(Q)$ gravity models. Finally, we offer concluding remarks and directions for future research in section \ref{remarks}.\\

 \section{$f(Q)$ framework} \label{$f(Q)$ framework}

The primary emphasis of this research lies in $f(Q)$ modified gravity models, distinguished by the presence of the non-metricity tensor \cite{Jimnez,FF}
\begin{equation}
Q_{\alpha \mu \nu}=\nabla_{\alpha }g_{\mu \nu } ,
\end{equation}
where $g_{\mu \nu }$ is the metric tensor. The non-metricity scalar $Q$ is expressed as
\begin{equation}
Q=-Q_{\alpha \mu \nu}P^{\alpha \mu \nu}.
\label{Q}
\end{equation}
The tensor $P^{\alpha \mu \nu}$ represents the non-metricity conjugate
\begin{equation}
P^{\alpha}_{ \mu \nu}=-\frac{1}{2}L_{\mu \nu }^{\alpha }+\frac{1}{4}\left ( Q^{\alpha }-\tilde{Q}^{\alpha } \right )g_{\mu \nu }-\frac{1}{4}\delta^{\alpha}_{(\mu}Q_{\nu)},
\end{equation}
with
\begin{equation}
L_{\mu \nu }^{\alpha }=\frac{1}{2}Q_{\mu \nu }^{\alpha }-Q_{(\mu \nu )}^{\alpha },
\end{equation}
and
\begin{equation}
\begin{array}{cc}
Q_{\alpha}=g^{\mu \nu}Q_{\alpha \mu \nu}, & \qquad \tilde{Q}_{\alpha}=g^{\mu \nu}Q_{\mu\alpha \nu}.
\end{array}
\end{equation}

The $f(Q)$ modified gravity model under study is based on the following action \cite{JB}

\begin{equation}
\mathit{S}=\int d^{4}x\sqrt{-g}\left [ -\frac{f(Q)}{16\pi G}+\mathfrak{\Lb_{m}} \right ].
\label{S}
\end{equation}

In this expression, $\mathfrak{\Lb_{m}}$ denotes the Lagrangian for matter fields, $G$ represents the Newtonian constant, $f(Q)$ is a general function of the non-metricity scalar, and $g$ stands for the determinant of the metric tensor $g_{\alpha \beta}$. In flat space-time, the action (\ref{S}) is equivalent to General Relativity when $f(Q)=Q$  \cite{JB}. The energy-momentum tensor can be represented as 

\begin{equation}
T_{\mu \nu }=\frac{-2}{\sqrt{-g}}\frac{\delta \left ( \sqrt{-g}\mathfrak{\Lb_{m}} \right )}{\delta g^{\mu \nu }}.
\end{equation}

Through the variation of the modified Einstein-Hilbert action (\ref{S}) concerning the metric tensor $g_{\alpha \beta}$, the gravitational field equations can be represented as follows

\begin{widetext} 
\begin{equation}
\frac{-2}{\sqrt{-g}}\Delta_{\alpha }(\sqrt{-g}f_{Q}P_{\mu \nu }^{\alpha })-\frac{1}{2}g_{\mu \nu }f-f_{Q}(P_{\mu \alpha i}Q_{\nu }^{\alpha i}-2Q_{\alpha i\mu }P^{\alpha i}_{\nu })
=T_{\mu \nu } ,
\end{equation}
\end{widetext}
here, $f_{Q}= \frac{\partial f}{\partial Q}$. The variation of the action (\ref{S}) can be expressed as

\begin{equation}
\nabla_{\mu }\nabla_{\nu}(\sqrt{-g}f_{Q}P^{\mu \nu }_{\alpha })=0 .
\end{equation}
To integrate this gravitational theory into cosmology, we consider the flat Friedmann-Lemaître-Robertson-Walker (FLRW) line element, which is defined as
\begin{equation}
ds^2=-d t^2+a^2(t)\left[d r^2+r^2\left(d \theta^2+\sin ^2 \theta d \phi^2\right)\right],
\label{ds}
\end{equation}
where $a(t)$ and $t$ represent the scale factor, and the cosmic time, respectively.
The non-metricity invariant $Q$ as specified in Eq.(\ref{Q}), is expressed as $Q = 6H^{2}$, where $H=\frac{\dot{a}}{a}$ represents the Hubble function, and the dot indicates the derivative with respect to the cosmic time $t$. We suppose that the energy-momentum tensor is represented by a perfect fluid, i.e.
\begin{equation}
T_{\mu \nu }=\left ( p+\rho  \right )u_{\mu}u_{\nu }+P g_{\mu \nu } ,
\label{T}
\end{equation}
here, $P$ stands for pressure, and $\rho$ represents energy density.\\
Using Eqs (\ref{ds}) and (\ref{T}), we deduce the modified Friedmann equations for $f(Q)$ gravity \cite{JB,Jimnez}
\begin{equation}
6H^{2}f_{Q}-\frac{1}{2}f=8 \pi G \rho ,
\label{eq1}
\end{equation}
\begin{equation}
\left ( 12H^{2}f_{QQ}+f_{Q} \right )\dot{H}=-4 \pi G \left ( 
 \rho+P \right ),
\label{eq2}
\end{equation}
where $f_{QQ}= \frac{\partial^{2} f}{\partial Q^{2}}$. \\

In this study, we introduce a cosmological form of $f(Q)$ gravity as motivated in \cite{Capozziello,Mhamdi2}

\begin{equation}
f(Q)=\alpha +\beta Q^{n},
\label{f1}
\end{equation}        
where $\alpha$, $\beta$, and $n$ represent the free parameters of the model. This model possesses intriguing characteristics that establish connections with both General Relativity and $\Lambda$CDM. More precisely, when $\alpha = 0$ and $\beta = n = 1$, the model aligns with General Relativity. Conversely, for $n = \beta = 1$ and $\alpha > 0$, it reflects the $\Lambda$CDM model.\\

We narrow our focus to the scenario where matter dominates i.e. $\rho=\rho_{m}$ and $P=P_{m}=0$, representing a Universe consisting solely of matter without the presence of dark energy or radiation. In this context, using Eqs. (\ref{eq1}), (\ref{eq2}) and (\ref{f1}), the dynamical equation that describes our model can be written as follow
\begin{equation}
\dot{H}=-\frac{3 H^2}{2n}\left(\frac{\alpha  6^{-n} H^{-2n}}{\beta -2 \beta  n}+1\right).
\label{dyna}
\end{equation}
The dynamics, described by Eq. (\ref{dyna}), can be reformulated into the following expression
\begin{equation}
\frac{dH}{dz}-\frac{3 H }{2 n (z+1)}\left(\frac{\alpha  6^{-n} H^{-2n}}{\beta -2 \beta  n}+1\right)=0,
\label{dH}
\end{equation}
where we have used the following conversion
\begin{equation}
\frac{dH}{dt}=-\left ( 1+z \right) H(z)\frac{dH}{dz}.
\end{equation}
By employing the current value of the Hubble parameter $H(z =0)= H_{0}$, and the normalized Hubble parameter, represented as $E^{2}\left ( z \right )=H^{2}(z)/ H^{2}_{0}$, we deduce the following subsequent solution of the differential equation (\ref{dH}), as follows

\begin{equation}
E^{2}\left ( z \right )=\left [ \left(\frac{6^{-n} 3 \Omega _{\alpha }}{\beta -2 \beta  n}+1\right)(z+1)^3-\frac{6^{-n} 3 \Omega _{\alpha }}{\beta -2 \beta  n} \right ]^{1/n},
\label{EE}
\end{equation}

here, $\Omega _{\alpha }$ denotes a dimensionless quantity dependent on $\alpha$, $n$, and $H_{0}$, given by $\Omega {\alpha }=\frac{\alpha }{3H_{0}^{2n}}$. When $n=1$, and by analogy with the standard model, we observe that the dimensionless energy density of matter and dark energy are equal to $\Omega _{m}=1-\frac{\Omega _{\alpha }}{2\beta }$ and $\Omega _{\Lambda}=\frac{\Omega _{\alpha }}{2\beta }$, respectively.

\section{Linear matter perturbations}\label{matter perturbations}
In this section, we delve into the main points of the linear growth of matter fluctuations, exploring the incorporation of the growth index, $\gamma$, into the current analysis.
Within the framework of any model concerning dark energy, including alternative theories such as modified gravity, it is widely acknowledged that at subhorizon scales, the dark energy component is expected to be smooth. Consequently, we can consider perturbations solely within the matter component of the cosmic fluid \cite{fluid1}. On sub-horizon scales, the differential equation governing linear matter perturbations is expressed as \cite{fr,fluid3,fluid4,fluid5}
\begin{equation}
\ddot{\delta}_m+2 \nu H \dot{\delta}_m-4 \pi G_{\text{eff}} \rho_m \delta_m=0.
\label{delta}
\end{equation}
Where $G_{\text{eff}} = G_N \mu(t)$, with $G_N$ representing Newton's gravitational constant, while $\rho_{\mathrm{m}} \propto a^{-3}$ represents the the matter density. In summary, for standard General Relativity, $\mu  = 1$, whereas in modified gravity models, $\mu \neq 1$. The solution of the aforementioned equation (\ref{delta}) is given by $\delta_m \propto D(a)$, where $D(a)$ represents the growth factor. Regardless of the gravity type, the growth rate and the growth factor of clustering are expressed through the following parameterization \cite{f(a)}
\begin{equation}
f(a)=\frac{d \ln \delta_m}{d \ln a} \simeq \Omega_m^\gamma(a),
\label{f}
\end{equation}
and,
\begin{equation}
D(a)=\exp \left[\int_1^{a(z)} \frac{\Omega_m^\gamma(y)}{y} d y\right],
\label{d}
\end{equation}
respectively. In this context, $\Omega_m(a) = \frac{\Omega_{m0}}{E^2(a)} a^{-3}$, and $\gamma$ represents the growth index. The growth index plays a crucial role as it serves as a distinguishing factor between general relativity and modified gravity at cosmological scales. For instance, in the case of a constant equation of state (EoS) parameter of dark energy, $\omega_{de}$, the growth index within the framework of general relativity is approximated by $\gamma \simeq \frac{3(\omega_{de}-1)}{6 \omega_{de}-5}$ \cite{Lue, Silveira,Wang,Nesseris} which reduces to $6/11$ for the concordance $\Lambda$CDM cosmology $(\omega_{de} = -1)$. However, for $f(R)$ theory the growth index parameter takes the range of  $0.40 \lesssim \gamma \lesssim 0.43$ \cite{fluid3,Tsujikawa}. By substituting the operator $\frac{d}{dt} = H \frac{d}{d\ln a}$ and Eq. (\ref{f}) into Eq. (\ref{delta}), we obtain the following expression
\begin{equation}
\frac{d f}{d \ln a}+f^2+\left(\frac{\dot{H}}{H^2}+2\right) f=\frac{3}{2} \mu(a) \Omega_m(a).
\label{ff}
\end{equation}
In the context of the dark energy model with an EoS parameter, $\omega_{de}$, using Friedmann equations
\begin{equation}
	\Big(\frac{\dot{a}}{a}\Big)^2=\frac{8\pi G}{3} (\rho_m+\rho_{de}),
\label{fried1}
\end{equation}
\begin{equation}
2\frac{\ddot{a}}{a}+\Big(\frac{\dot{a}}{a}\Big)^2=-{8\pi G}  p_{de},
\label{fried2}
\end{equation}
and setting
\begin{equation}
\Omega=\frac{8\pi G\rho_m}{3H^2}=\frac{\rho_m}{\rho_m+\rho_{de}},
\end{equation}
one can show that
\begin{equation}
\frac{\dot{H}}{H^2}+2=\frac{1}{2}-\frac{3}{2} \omega_{de}(a)\left[1-\Omega_{\mathrm{m}}(a)\right].
\end{equation}
%Where $\omega_{de}(a) = -1$, in this case of \lambda, the Hubble parameter is given by $H(a) = H_0 \sqrt{\Omega_{m0} a^{-3} + \Omega_{\Lambda0}}$. Here, $H_0$ represents the Hubble constant in the present epoch, $\Omega_{m0}$ is the present-day density parameter for matter, and $\Omega_{\Lambda0}$ is the present-day density parameter for dark energy.\\
In general, the growth index may exhibit redshift dependence, $\gamma(z)$, rather than remaining constant, $\gamma$. In this context, the insertion of Eq. (\ref{f}) into Eq. (\ref{ff}) and using
\begin{equation}
d\Omega=3\Omega\omega_{de}\Big(1-\Omega\Big)dx,
\end{equation}
one obtains
%\begin{equation}
 %-(1+z) \gamma^{\prime} \ln \left(\Omega_m\right)+\Omega_m^\gamma+3 \omega_{de}\left(1-\Omega_m\right)\left(\gamma-\frac{1}{2}\right) +\frac{1}{2}=\frac{3}{2}  \mu \Omega_m^{1-\gamma},
%\label{gamma111}
%\end{equation}

\begin{equation}
\begin{split}
&-(1+z) \gamma^{\prime} \ln \left(\Omega_m\right)+\Omega_m^\gamma + \\
&\quad\quad 3 \omega_{de}\left(1-\Omega_m\right)\left(\gamma-\frac{1}{2}\right)\frac{1}{2}=\frac{3}{2}  \mu \Omega_m^{1-\gamma},
\end{split}
\label{gamma111}
\end{equation}
where the prime signifies the derivative with respect to redshift $z$. Evaluating Eq (\ref{gamma111}) at $z=0$ we have
%\begin{equation}
%-\gamma^{\prime}(0) \ln \left(\Omega_{m 0}\right)+\Omega_{m 0}^{\gamma(0)}+3 \omega_{de0}\left(1-\Omega_{m 0}\right)\left[\gamma(0)-\frac{1}{2}\right]+\frac{1}{2} =\frac{3}{2} \mu_0 \Omega_{m 0}^{1-\gamma(0)},
%\label{Gamma111}
%\end{equation}

\begin{equation}
\begin{split}
&-\gamma^{\prime}(0) \ln \left(\Omega_{m 0}\right)+\Omega_{m 0}^{\gamma(0)}+3 \omega_{de0}\left(1-\Omega_{m 0}\right)\left[\gamma(0)-\frac{1}{2}\right]\\
&\quad\quad\quad\quad\quad\quad\quad\quad\quad\quad\quad\quad\quad+\frac{1}{2} =\frac{3}{2} \mu_0 \Omega_{m 0}^{1-\gamma(0)}.
\end{split}
\label{Gamma111}
\end{equation}

In our previous work \cite{Mhamdi2}, we have constrained the $f(Q)$ model by solving numerically $\delta_{m}(a)$  from the differential equation given in Eq. (\ref{delta}) where $f(a)=\frac{d \delta(a)}{d \ln a}$. However,  in this current study, the expression of growth factor $ f$ becomes described in Eq. (\ref{f}). Here, we explore four distinct functional forms of $\gamma(z)$, where $\gamma(z)=\gamma_0+y(z)\gamma_{1}$, with the expression of $y(z)$ will be specified bellow. These functional forms represent a first-order Taylor expansion around specific cosmological quantities such as the scale factor and redshift. These parameterizations are defined as follows \cite{f(a),Polarski,Belloso,Ishak,DiPorto,Basilakos1}

\begin{equation}
\gamma(z) = \begin{cases}
     \gamma_{0}, & \Gamma_0 \text{-parametrization} \\
       \gamma_{0} + z   \gamma_{1}, &  \Gamma_1 \text{-parametrization} \\
      \gamma_{0} + \frac{z}{1+z}  \gamma_{1}, &  \Gamma_2 \text{-parametrization} \\
         \gamma_{0}+ z e^{-z}  \gamma_{1}, &  \Gamma_3 \text{-parametrization}
\end{cases}
\label{para0-3}
\end{equation}

The $\Gamma_0$ model represents a constant growth index \cite{f(a)}, with $\gamma_{1}$ strictly set to zero, thus $\gamma = \gamma_0$. The $\Gamma_1$ model represents an expansion around $z = 0$ \cite{Polarski}, where $y(z) = z$. However, this parametrization is valid only at relatively low redshifts, $0 \leq z \leq 0.5$. The $\Gamma_2$ model represents an expansion around $a=1$ \cite{Belloso,Ishak,DiPorto}, where the function $y$ becomes $y(z)=1-a(z)=1-\frac{1}{1+z}$. At large redshifts $z \gg 1$, we have $\gamma_{\infty} \simeq \gamma_0+\gamma_1$. The last model $\Gamma_3$ represents an interpolated parametrization, with the function $y$ reformulated as $y(z) = ze^{-z}$ \cite{Basilakos1}, which smoothly connects low and high-redshift ranges. Clearly, at large redshifts $z \gg 1$, we have $\gamma_{\infty} \simeq \gamma_0$.\\

Obviously, using the $\Gamma_{1-3}$-parametrization mentioned above, which satisfies $y(z=0)=0$ and $\left.\gamma^{\prime}(0)=\gamma_1 y^{\prime}(0)\right)$, and by using Eq. (\ref{Gamma111}), we can express the parameter $\gamma_1$ in relation to $\gamma_0$ as follows:

\begin{equation}
\gamma_1=\frac{\Omega_{m 0}^{\gamma_0}+3 \omega_{de 0}\left(\gamma_0-\frac{1}{2}\right)\left(1-\Omega_{m 0}\right)-\frac{3}{2} \mu_0 \Omega_{m 0}^{1-\gamma_0}+\frac{1}{2}}{y^{\prime}(0)\ln \Omega_{m 0}} .
\label{G1}
\end{equation}

From the above forms of $\gamma(z)$ it becomes evident that for the $\Gamma_{1-3}$ parametrizations, we have $y(0)=0$ and $y^{\prime}(0)=1$. Therefore, for the case of the $\Lambda$CDM model with $\left(\Omega_{m 0}, \gamma_0\right)=\left(0.273, \frac{6}{11}\right)$, Eq. (\ref{G1}) provides $\gamma_1 \simeq-0.0478$.\\

%In our previous research \cite{Mhamdi2}, we have constrained the $f(Q)$ model Eq. (\ref{EE}) by solving numerically $\delta$ from Eq. (\ref{delta}), However, in this study, we use the parameterization outlined in Eq. (\ref{f}), where the growth index $\gamma$ being as a free parameter.

 \section{Statistical analysis} \label{statistics}

In this section, our attention is directed towards constraining the growth index parameters presented in the previous section within each $f(Q)$ parameterization models $\Gamma_{0-3}$, where $\gamma_0$, $\gamma_1$, and $\sigma_{8}$ are treated as a free parameter. We aim to constrain their value using the latest cosmological data, encompassing both background and perturbation levels. To achieve this, we employ a statistical technique known as Markov Chain Monte Carlo (MCMC) analysis \cite{mcmc1}. The latter, constrains the model  by varying them in a range of conservative priors and exploring the posteriors of the parameter space \cite{Gilks}. Therefore, for each parameter, we obtain its one- and two-dimensional distributions, where the one-dimensional distribution represents the parameters’ posterior distribution whilst the two-dimensional one illustrates the covariance between two different parameters. We employ maximum likelihood estimation for fitting, using Python GetDist package \cite{python}.
This method maximizes the total likelihood function, $\mathcal{L}_{\text{tot}}$, by minimizing $\chi_{\text{tot}}^2$. We use several observational datasets: type Ia supernova data (Pantheon$^{+}$, 1701 points), cosmic chronometer data ($\mathrm{CC}$, 57 points), and redshift space distortion data (RSD, 20 points).

\begin{table*}[t!]
\centering
\begin{tabular}{|c|c|c|c|c|c|c|c||c|c|c|c|c|c|c|c|}
\hline 
\multicolumn{8}{|c||}{ DA method } & \multicolumn{8}{c|}{ BAO method } \\
\hline
 $z$ & $H(z)$ & $\sigma_H$ & Refs & $z$ & $H(z)$ & $\sigma_H$ & Refs & $z$ & $H(z)$ & $\sigma_H$ & Refs & $z$ & $H(z)$ & $\sigma_H$ & Refs \\
\hline
0.070 & 69 & 19.6 & \cite{42} &  0.4783 & 80.9 & 9 &  \cite{45}   &0.24 & 79.69 & 2.99 & \cite{34}  & 0.57 & 96.8 & 3.4    & \cite{32}  \\
0.090 & 69 & 12 & \cite{40}&     0.480 & 97 & 62 & \cite{41}        & 0.30 & 81.7  & 6.22 & \cite{red}  & 0.59 & 98.48 & 3.18 & \cite{33}  \\
0.120 & 68.6 & 26.2 & \cite{42} &  0.593 & 104 & 13 & \cite{43}         & 0.31 & 78.18 & 4.74 & \cite{33}   & 0.60 & 87.9 & 6.1 & \cite{35}  \\
0.170 & 83 & 8 & \cite{40}  & 0.6797 & 92 & 8 & \cite{43}              & 0.34 & 83.8 & 3.66 & \cite{34}   & 0.61 & 97.3 & 2.1 & \cite{38}  \\
0.1791 & 75 & 4 & \cite{43}  & 0.7812 & 105 & 12 & \cite{43}        & 0.35 & 82.7 & 9.1 & \cite{28}  & 0.64 & 98.82 & 2.98 & \cite{33}   \\
0.1993 & 75 & 5 & \cite{43}  & 0.8754 & 125 & 17 & \cite{43}       & 0.36 & 79.94 & 3.38 & \cite{33}  & 0.73 & 97.3 & 7.0 & \cite{35}  \\
0.200 & 72.9 & 29.6 & \cite{42}&  0.880 & 90 & 40 & \cite{41}      & 0.38 & 81.5 & 1.9 & \cite{38}      & 2.30 & 224 & 8.6 & \cite{36}   \\
0.270 & 77 & 14 & \cite{40}       & 0.900 & 117 & 23 & \cite{40}     & 0.40 & 82.04 & 2.03 & \cite{33}   & 2.33 & 224 & 8 & \cite{39}   \\
0.280 & 88.8 & 36.6 & \cite{42}& 1.037 & 154 & 20 & \cite{43}      & 0.43 & 86.45 & 3.97 & \cite{34}   & 2.34 & 222 & 8.5 & \cite{31}  \\
0.3519 & 83 & 14 & \cite{43}  & 1.300 & 168 & 17 & \cite{40}         & 0.44 & 82.6 & 7.8 & \cite{35}   & 2.36 & 226 & 9.3 & \cite{30}  \\
0.3802 & 83 & 13.5 & \cite{45}  & 1.363 & 160 & 33.6 & \cite{44}     & 0.44 & 84.81 & 1.83 & \cite{33}        & & & & \\
0.400 & 95 & 17 & \cite{40}  &  1.430 & 177 & 18 & \cite{40}          & 0.48 & 87.79 & 2.03 & \cite{33}       & & & & \\
0.4004 & 77 & 10.2 & \cite{45}   &  1.530 & 140 & 14 & \cite{43}        & 0.51 & 90.4 & 1.9 & \cite{38}          & & & &  \\
0.4247 & 87.1 & 11.2 & \cite{45}   & 1.750 & 202 & 40 &\cite{43}       & 0.52 & 94.35 & 2.64 & \cite{33}     & & & & \\
0.4497 & 92.8 & 12.9 & \cite{45}    & 1.965 & 186.5 & 50.4 & \cite{44}   & 0.56 & 93.34 & 2.3 & \cite{33}     & & & &  \\
0.470 & 89 & 34 & \cite{46}  &       & & &                                    & 0.57 & 87.6 & 7.8 & \cite{29}         & & & & \\

\hline
\end{tabular}
\caption{Hubble parameter values $H(z)$ with errors $\sigma_H$ from DA and BAO methods.}

\label{CCdata}
\end{table*}
\begin{table*}[t!]
\centering
\begin{tabular}{|c|c|c|c|c|}
\hline
$z$ & $f \sigma_8$ & \text{Survey} & \text{Cosmological Tracer} & \text{Reference} \\
\hline
0.02 & $0.398 \pm 0.065$ & \text{SnIa+IRAS} & \text{SNIa + galaxies} & \cite{f42} \\
0.025 & $0.39 \pm 0.11$ & \text{6dFGS} & \text{voids} & \cite{f43} \\
0.067 & $0.423 \pm 0.055$ & \text{6dFGS} & \text{galaxies} & \cite{f44} \\
0.10 & $0.37 \pm 0.13$ & \text{SDSS-veloc} & \text{DR7 galaxies} & \cite{f45} \\
0.15 & $0.53 \pm 0.16$ & \text{SDSS-IV} & \text{eBOSS DR16 MGS} & \cite{f46} \\
0.32 & $0.384 \pm 0.095$ & \text{BOSS-LOWZ} & \text{DR10, DR11} & \cite{f47} \\
0.38 & $0.497 \pm 0.045$ & \text{SDSS-IV} & \text{eBOSS DR16 galaxies} & \cite{f46} \\
0.44 & $0.413 \pm 0.080$ & \text{WiggleZ} & \text{bright emission-line galaxies} & \cite{f48} \\
0.57 & $0.453 \pm 0.022$ & \text{CMASS-BOSS} & \text{DR12 voids+galaxies} & \cite{f49} \\
0.59 & $0.488 \pm 0.060$ & \text{SDSS-CMASS} & \text{DR12} & \cite{f50} \\
0.70 & $0.473 \pm 0.041$ & \text{SDSS-IV} & \text{eBOSS DR16 LRG} & \cite{f46} \\
0.73 & $0.437 \pm 0.072$ & \text{WiggleZ} & \text{bright emission-line galaxies} & \cite{f48} \\
0.74 & $0.50 \pm 0.11$ & \text{SDSS-IV} & \text{eBOSS DR16 voids} & \cite{f2} \\
0.76 & $0.440 \pm 0.040$ & \text{VIPERS v7} & \text{galaxies} & \cite{f51} \\
0.85 & $0.52 \pm 0.10$ & \text{SDSS-IV} & \text{eBOSS DR16 voids} & \cite{f2} \\
0.978 & $0.379 \pm 0.176$ & \text{SDSS-IV} & \text{eBOSS DR14 quasars} & \cite{f52} \\
1.05 & $0.280 \pm 0.080$ & \text{VIPERS v7} & \text{galaxies} & \cite{f51} \\
1.40 & $0.482 \pm 0.116$ & \text{FastSound} & \text{ELG} & \cite{f53} \\
1.48 & $0.30 \pm 0.13$ & \text{SDSS-IV} & \text{eBOSS DR16 voids} & \cite{f2} \\
1.944 & $0.364 \pm 0.106$ & \text{SDSS-IV} & \text{eBOSS DR14 quasars} & \cite{f52} \\
\hline
\end{tabular}
\label{Growthdata}
\caption{Data compilation of 20 $f \sigma_8(z)$ measurements.}
\end{table*}

\subsection{Background data}

\begin{itemize}
\item \textbf{Pantheon$^{+}$:} We include the updated Pantheon$^+$ compilation supernovae dataset \cite{SNdata} which represents a significant advancement in our understanding of the universe’s expansion history. These datasets, made of 1701 data points, are obtained from 1550 Type Ia supernovae spanning a redshift range of $0.001\leq  z\leq  2.3$. The $\chi^{2}$ function for the supernovae data sets is given by

\begin{equation}
\chi_{\text{Pantheon}^+}^{2}\left(P_Q, M, H_0\right)=\vec{D}^T \cdot\left(C_{\text {stat }+ \text { sys }}\right)^{-1} \cdot \vec{D},
\label{CHISNIA}
\end{equation}

where $\vec{D}$ is a 1701-dimensional vector and $C_{\text {stat+sys}}$ represents the covariance matrix obtained from the Pantheon$^+$ sample accounting for both statistical and systematic uncertainties.
The free parameters of the model are $P_Q, M,$ and $H_0$ where $P_Q$ represents the vector of parameters $P_{Q}=\left \{ \Omega_{\alpha},\beta,n \right \}$ of $f(Q)$ model, and $M$ represents the absolute magnitude. The Pantheon+ sample has $1701 \mathrm{SN}$ light curves, 77 of which correspond to galaxies hosting Cepheids in the low redshift range $0.00122 \leq z \leq 0.01682$. In order to break the degeneracy between $H_0$ and the absolute magnitude $M$ of Type Ia, we define the vector
$$
D^{\prime}_i=\left\{\begin{array}{l}
m_i-M-\mu_i^{\text {Ceph }}, \quad  \in \text { Cepheid hosts } \\
m_i-M-\mu_{\text {model }}\left(z_i\right), \quad \text { otherwise }
\end{array}\right.
$$
where $\mu_i^{\text {Ceph }}$ denotes the distance modulus associated with the Cepheid host of the $i^{\text {th}}$ SNIa, independently measured through Cepheid calibrators. $m_{Bi}$ and $\mu_{model}$ represent the apparent magnitudes and the predicted distance modulus, respectively. The predicted distance modulus is given according to a selected cosmological model, as follows
%\begin{align}
 %  & \mu_{\text{model}}(z_{i},\Omega_{\alpha},\beta,n,H_{0}) = \notag \\
  %  &\quad 5\log_{10}D_{L}(z_{i},\Omega_{\alpha},\beta,n,H_{0}) + 25,
%\end{align}
\begin{equation}
\mu_{\text{model}} (z_{i},P_Q, H_0)= 5\log_{10}D_{L}\left(z_{i},P_Q, H_0\right) + 25,
\end{equation}
where $D_{L}$ denotes the luminosity distance, defined as
%\begin{align}
%&D_{L}\left(z_{i},\Omega_{\alpha},\beta,n\right,H_{0}) = \notag \\
%&\quad  \left(1+z\right)\int_{0}^{z}\frac{c dz'}{H\left(z',\Omega_{\alpha},\beta,n,H_{0}\right)} ,
%\end{align}
\begin{equation}
D_{L}(z_{i},P_Q, H_0) = \left(1+z\right)\int_{0}^{z}\frac{c dz'}{H\left(z',P_Q, H_0 \right)} .
\end{equation}
Consequently, Eq. (\ref{CHISNIA}) can be rewritten as
\begin{equation}
\chi_{\mathrm{SN}}^2=\vec{D}^{\prime T} \cdot \mathbf{C}_{\text {Pantheon }^+}^{-1} \cdot \vec{D}^{\prime}
\end{equation}
\end{itemize}
\begin{itemize}
\item \textbf{Cosmic Chronometer data:} In Cosmic Chronometry, the Hubble parameter $H$ values at specific redshifts $z$ can be determined through two approaches: (i) by extracting $H(z)$ from line-of-sight BAO data \cite{BAOcc,BAOcc1} including analysis of correlation functions of luminous red galaxies \cite{BAOcc,red}, and (ii)  by estimating $H(z)$ from differential ages $\Delta t$ of galaxies using DA method \cite{DAcc,DAcc1}, which relies on the following relation

\begin{equation}
H(z)=-\frac{1}{1+z} \frac{d z}{d t} \simeq-\frac{1}{1+z} \frac{\Delta z}{\Delta t}.
\end{equation}
The maximal set with 57 data points of $H(z)$ measurements spanning the redshift range $0 < z \leq  2.36$ is shown in Table \ref{CCdata} below, it includes 31 data points measured with DA method and 26 data points, obtained with $\mathrm{BAO}$. The chi-square value for the cosmic chronometers is calculated according to the following definition
\begin{equation}
\chi^{2}_{CC}(P_Q, H_0)=\sum_{i=1}^{57}\left [ \frac{H_{obs}\left ( z_{i} \right )-H_{th}\left ( z_{i},P_Q, H_0 \right )}{\sigma_H \left ( z_{i} \right )} \right ]^{2} ,
\end{equation}
where $H_{\text{obs}}$ and  $H_{\text{th}}$ denote respectively the observed and the theoretical value of the Hubble parameter. On the other hand, $\sigma_H \left ( z_{i} \right )$ corresponds to the error on the observed values of the Hubble parameter $H(z)$. Throughout this paper, we denote this observation as ”CC”.
\end{itemize}

\subsection{Perturbation data}
\begin{itemize}
\item \textbf{Redshift space distortion:} One of the crucial probes of large scale structure is redshift space distortions, which provide measurements of $f \sigma_8(z)$. In galaxy redshift surveys, RSD measures the peculiar velocities of matter. The result is inferring the growth rate of cosmological perturbations. This measurement is done on a wide range of redshifts and scales. This can be obtained by measuring the ratio of the monopole and the quadrupole multipoles of the redshift space power spectrum, which depends on $\beta=f / b$. Here $f$ is the growth rate and $b$ is the bias the combination of $f \sigma_8(z)$ is independent of the bias factor and the bias dependence in this combination cancels out.  In this study, the growth data that we use is based on the 6dFGS, SDSS-veloc, SDSS-IV, BOSS-LOWZ, CMASS-BOS, SDSS-CMAS, VIPERS v7, FastSound, SnIa+IRAS,  and WiggleZ galaxy surveys (see  table \ref{Growthdata}). The $\chi^{2}_{\text{RSD}}$ function is defined as

\begin{equation}
\chi^{2}_{\text{RSD}}\left (P_{Q},P_{\gamma},\sigma _{8} \right )=\sum_{i=1}^{20}\left ( \frac{f\sigma_{8,i}-f\sigma_{8}(z_{i},P_{Q},P_{\gamma},\sigma_{8})}{\sigma(z_{i})} \right )^{2}.
\end{equation}

In this expression, $\sigma(z_{i})$ represents the standard error associated with the observed value of $f\sigma_{8}$. The terms $f\sigma_{8,i}$ and $f\sigma_{8}(z_i)$ correspond to the observed and theoretical values of $f\sigma_{8}$, respectively. The theoretical $f\sigma_{8}$ value is given by

\begin{equation}
f \sigma_8(z,P_{Q},P_{\gamma},\sigma_{8})=\sigma_8 \Omega_m(z)^{\gamma(z)} D(z,P_{Q},P_{\gamma},\sigma_{8}) ,
\end{equation}

here, $D(z,P_{Q},P_{\gamma},\sigma_{8})$ represents the growth factor as defined previously in Eq. (\ref{d}). The $P_{\gamma}$ represents the growth index parameters vector, which contains $\gamma_0$ and $\gamma_1$,  $P_{\gamma}=\left \{ \gamma_{0},\gamma_1\right \}$. The quantity $\sigma_{8,0}$ stands for the amplitude of the matter power spectrum at the scale of $8 h^{-1} Mpc$ at the present time, i.e. $z=0$. It is directly connected to the amplitude of the primordial fluctuations and is determined by the growth rate of cosmological fluctuations.

\end{itemize}

The total likelihood and chi-square functions are therefore given by
%\begin{equation}
%\mathcal{L}_{t o t}(P)=\mathcal{L}_{Pantheon^+}\left(P_{Q},H_0,M\right) \times \mathcal{L}_{CC}\left(P_{Q}, H_0\right) \times \mathcal{L}_{RSD}\left(P_{Q}, P_{\gamma}, %\sigma_8\right) 
%\end{equation}
\begin{equation}
\begin{split}
\mathcal{L}_{t o t}(P) =&  \mathcal{L}_{Pantheon^+}\left(P_{Q},H_0,M\right)  \times \mathcal{L}_{CC}\left(P_{Q}, H_0\right)\\
& \times \mathcal{L}_{RSD}\left(P_{Q}, P_{\gamma}, \sigma_8\right),
\end{split}
\end{equation}
and 
%\begin{equation}
%\chi_{tot}^{2}(P)=\chi^{2}_{Pantheon^+}\left(P_{Q},H_0,M\right) + \chi^{2} _{CC}\left(P_{Q}, H_0\right) + \chi^{2}_{RSD}\left(P_{Q}, P_{\gamma}, \sigma_8\right) 
%\end{equation}
\begin{equation}
\begin{split}
\chi_{tot}^{2}(P) = & \chi^{2}_{Pantheon^+}\left(P_{Q},H_0,M\right) + \chi^{2}_{CC}\left(P_{Q}, H_0\right)\\
& + \chi^{2}_{RSD}\left(P_{Q}, P_{\gamma}, \sigma_8\right),
\end{split}
\end{equation}
respectively. The vectors $P_{Q}$ and $P_{\gamma}$ contain the free parameters of the $f(Q)$ model. Nevertheless, the main parameters relevant to our study are $P_Q \equiv  (\Omega_{\alpha}, \beta, n)$ and $P_{\gamma} \equiv  (\gamma_{0}, \gamma_{1})$. Note that in the case of the $\Lambda$CDM we have $P_Q \equiv \Omega_m$. Table \ref{tab:priors} represents the priors of each parameter using in the MCMC analysis.
\begin{table}[h!]
    \centering
    \begin{tabular}{|c|c|}
        \hline
        \text{Parameter} & \text{Prior} \\
        \hline
        $\gamma_{0}$ & $\left [ 0,1 \right ]$ \\
        $\gamma_{1}$ & $\left [ -1,1 \right ]$\\
         $\sigma_8$ &  $\left [0,2  \right ]$\\
        $\Omega_{m}$ &  $\left [ 0,1 \right ]$\\
         $\Omega_{\alpha}$ & $\left [0,1  \right ]$ \\
          $\beta$ & $\left [ 0,1 \right ]$ \\
          $n$ & $\left [0,2  \right ]$ \\
          $H_0$ &  $\left [40,100  \right ]$ \\
         $M$  &     $\left [-20, -19 \right ]$ \\
        \hline
    \end{tabular}
    \caption{Prior imposed on different parameters.}
    \label{tab:priors}
\end{table}

\begin{table*}[t!]
    \centering
    \begin{tabular}{ccccccccccc}
       \toprule
        Model & Param & $\gamma_0$ & $\gamma_1$ & $\sigma_8$ & $\Omega_m$ & $\Omega_{\alpha}$ & $\beta$ & $n$ & $H_{0}$ & $M$ \\
        \midrule
        $\Lambda$CDM  & $\Gamma_0$ & $0.545_{-0.096}^{+0.096}$ & - & $0.808_{-0.0359}^{+0.0359}$ & $0.273_{-0.011}^{+0.011}$ & - & - & - & $70.228_{-0.661}^{+0.661}$ & $-19.363_{-0.018}^{+0.018}$ \\
                      & $\Gamma_1$ &  $0.547_{-0.096}^{+0.096}$ & $-0.0106_{-0.0014}^{+0.0014}$ & $0.8071_{-0.0349}^{+0.0349}$ & $0.2735_{-0.011}^{+0.011}$ & - & - & - & $70.222_{-0.674}^{+0.674}$ & $-19.363_{-0.018}^{+0.018}$ \\
                      & $\Gamma_2$ & $0.544_{-0.097}^{+0.097}$ & $-0.00827_{-0.00154}^{+0.00154}$ & $0.808_{-0.037}^{+0.037}$ & $0.273_{-0.011}^{+0.011}$ & - & - & - & $70.249_{-0.655}^{+0.655}$ & $-19.362_{-0.018}^{+0.018}$ \\
                      & $\Gamma_3$ & $0.542_{-0.092}^{+0.092}$ & $-0.047_{-0.007}^{+0.007}$ & $0.802_{-0.034}^{+0.034}$ & $0.273_{-0.011}^{+0.011}$ & - & - & - & $70.222_{-0.651}^{+0.651}$ & $-19.363_{-0.018}^{+0.018}$ \\ 
        \midrule
   $f(Q)$        & $\Gamma_0$ & $0.571_{-0.110}^{+0.095}$ & - & $0.762_{-0.037}^{+0.032}$ & - & $0.660_{-0.17}^{+0.24}$ & $0.348_{-0.11}^{+0.11}$ & $1.099_{-0.043}^{+0.043}$ & $69.708_{-0.697}^{+0.697}$ & $-19.371_{-0.018}^{+0.018}$ \\
                      & $\Gamma_1$ & $0.596_{-0.11}^{+0.097}$ & $-0.0527_{-0.0098}^{+0.0057}$ & $0.767_{-0.036}^{+0.03}$ & - & $0.72_{-0.12}^{+0.22}$ & $0.387_{-0.089}^{+0.11}$ & $1.093_{-0.042}^{+0.042}$ & $69.68_{-0.70}^{+0.70}$ & $-19.372_{-0.018}^{+0.018}$ \\
                      & $\Gamma_2$ & $0.61_{-0.11}^{+0.10}$ & $-0.22_{-0.12}^{+0.12}$ & $0.764_{-0.034}^{+0.030}$ & - & $0.758_{-0.091}^{+0.20}$ & $0.406_{-0.0075}^{+0.11}$ & $1.094_{-0.044}^{+0.039}$ & $69.70_{-0.70}^{+0.70}$ & $-19.371_{-0.018}^{+0.018}$ \\
                      & $\Gamma_3$ & $0.62_{-0.12}^{+0.10}$ & $-0.359_{-0.24}^{+0.093}$ &$0.760_{-0.033}^{+0.029}$&-& $0.74_{-0.11}^{+0.22}$& $0.391_{-0.078}^{+0.11}$ & $1.098_{-0.042}^{+0.042}$&  $69.68_{-0.69}^{+0.69}$  & $-19.372_{-0.018}^{+0.018}$ \\
        \bottomrule
    \end{tabular}
    \caption{Summary of the mean $\pm$ 1$\sigma$ values of the cosmological parameters using Pantheon$^{+}$+CC+RSD datasets for both $\Lambda$CDM and $f(Q)$ models.}
    \label{bf}
\end{table*}

\begin{figure*}[t]
    \centering
    \includegraphics[scale=0.32]{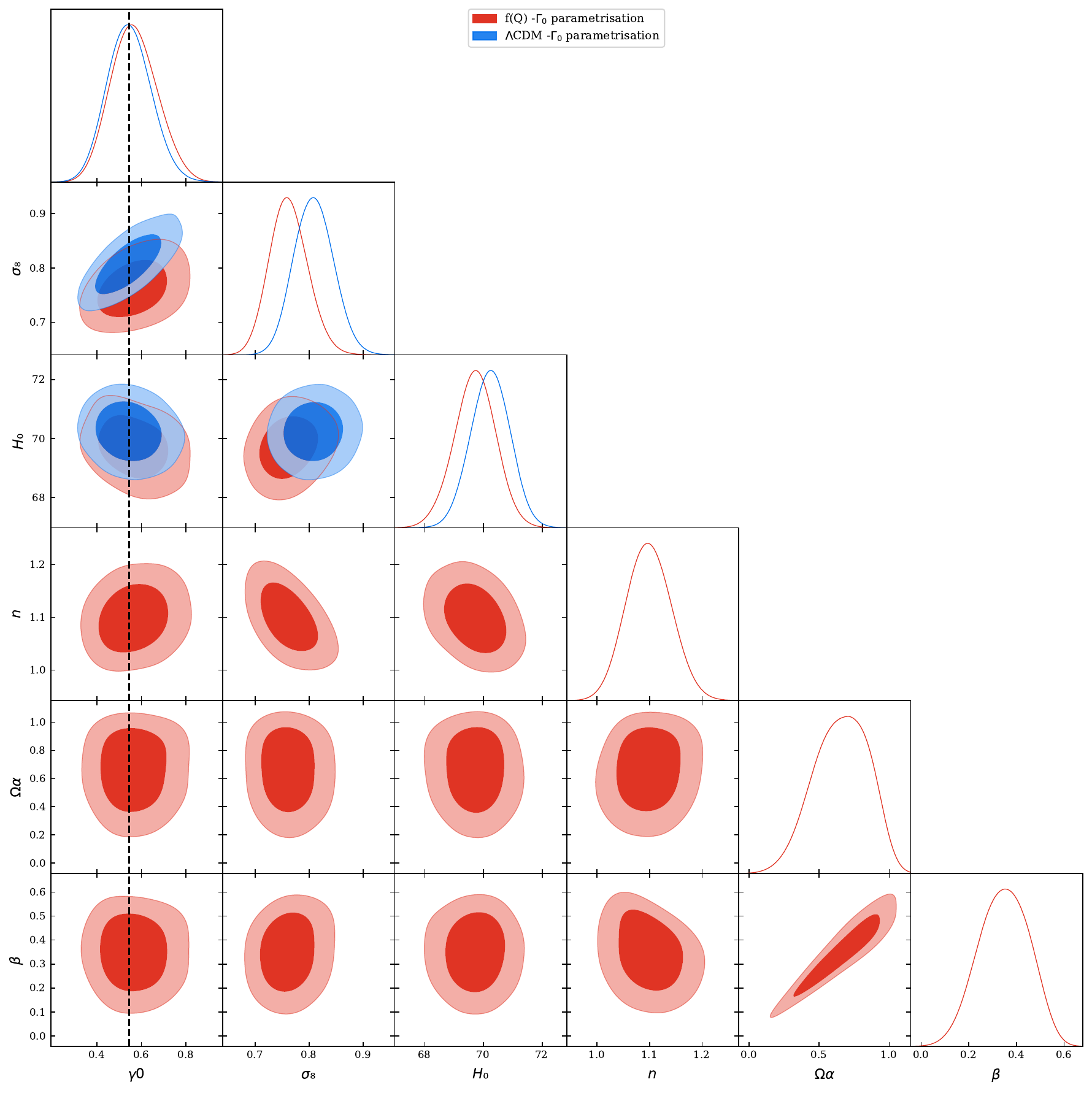}
    \caption{This figure shows the $1\sigma$ and $2\sigma$ confidence contours and the posterior distributions obtained from the RSD+CC+Pantheon$^+$ datasets for both $f(Q)$ and $\Lambda$CDM models using $\Gamma_0$ parametrization. The black dashed line represents the theoretical value of $\gamma_0=6/11$ for the $\Lambda$CDM model.}
\label{cont1}
\end{figure*}

\begin{figure*}[t]
    \centering
    \includegraphics[scale=0.32
]{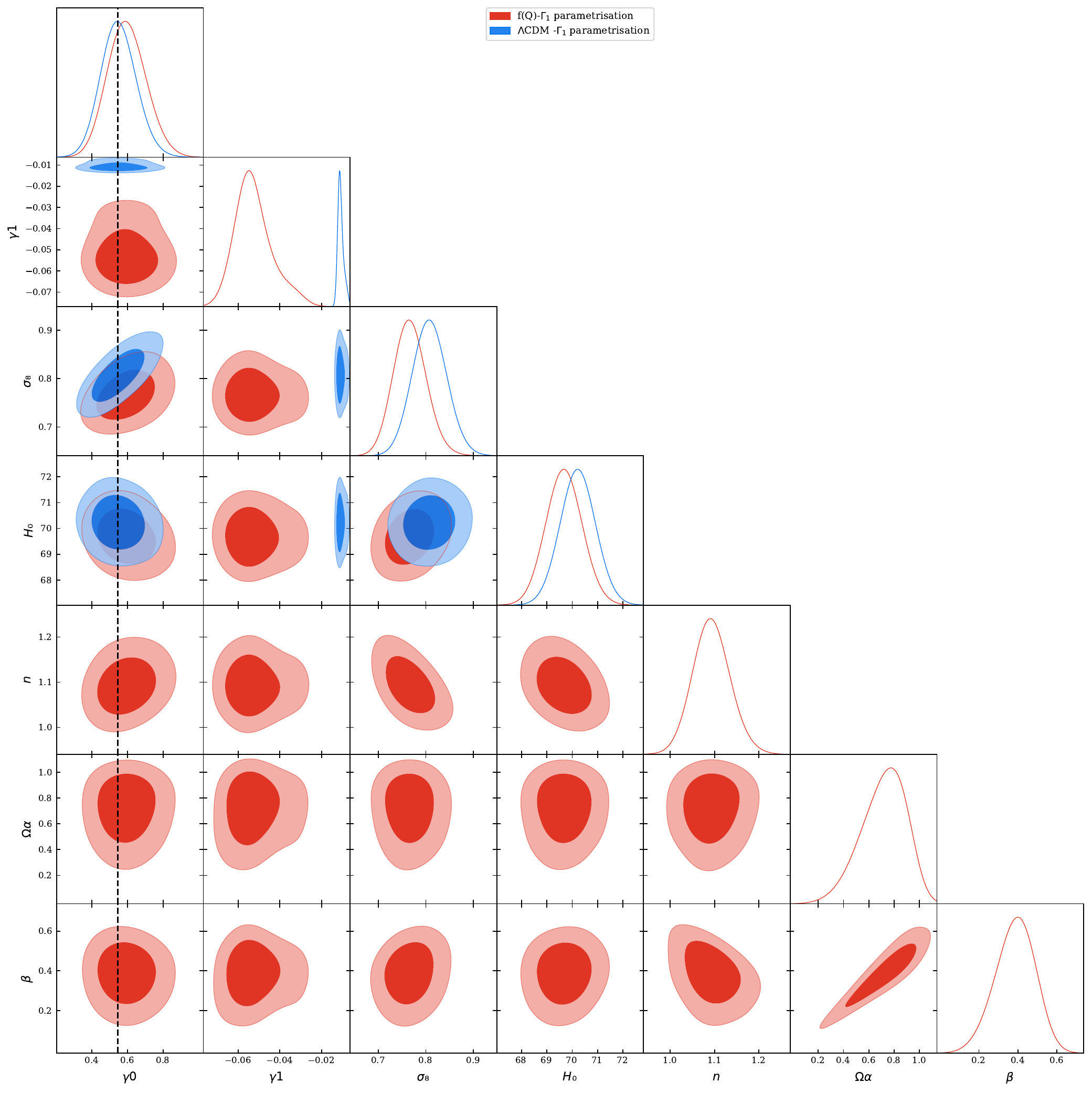}
    \caption{This figure shows the $1\sigma$ and $2\sigma$ confidence contours and the posterior distributions obtained from the RSD+CC+Pantheon$^+$ datasets for both $f(Q)$ and $\Lambda$CDM models usnig $\Gamma_1$ parametrization. The black dashed line represents the theoretical value of $\gamma_0=6/11$ for the $\Lambda$CDM model.}
\label{cont2}
\end{figure*}

\begin{figure*}[t]
    \centering
    \includegraphics[scale=0.32]{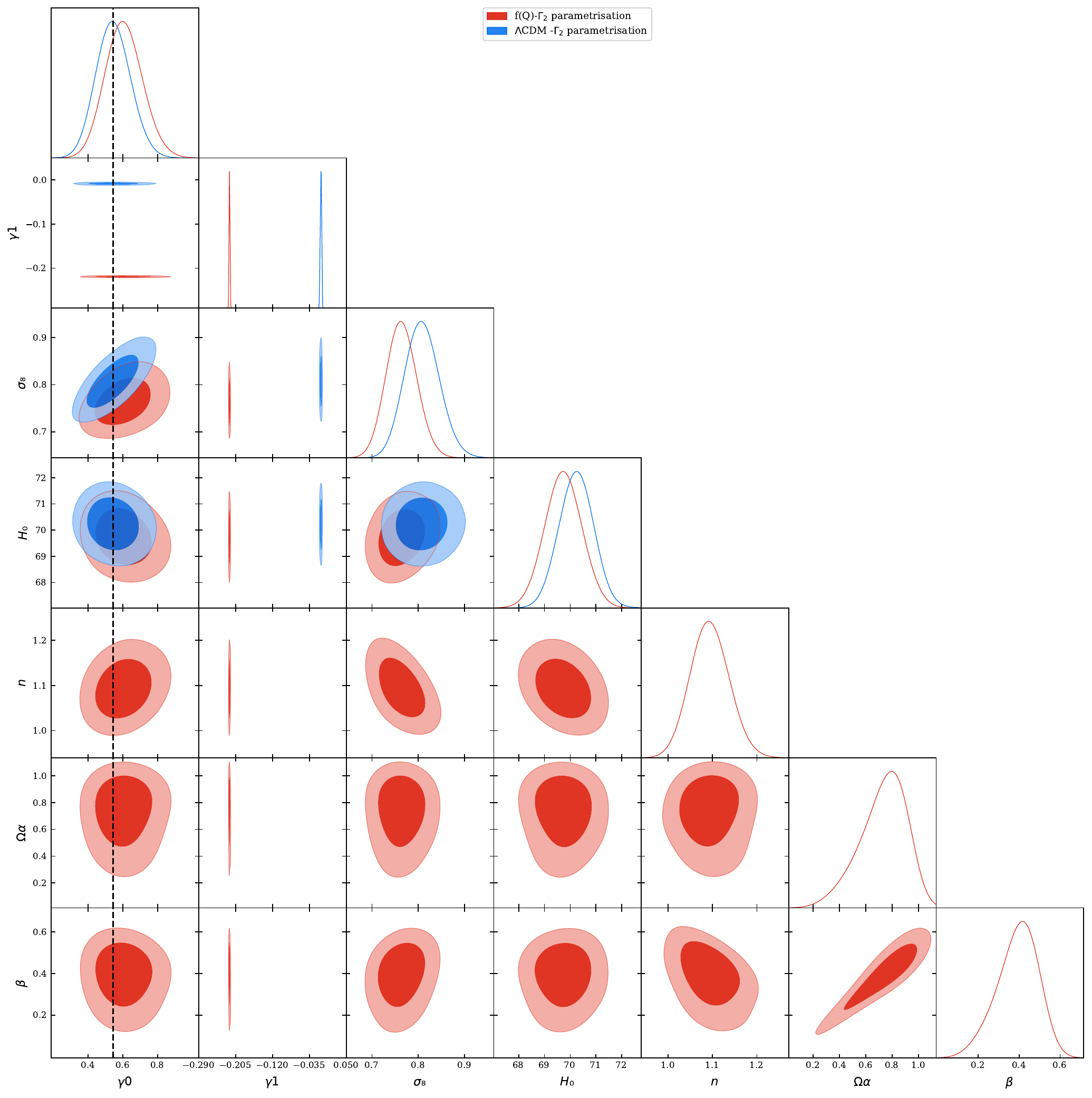}
    \caption{This figure shows the $1\sigma$ and $2\sigma$ confidence contours and the posterior distributions obtained from the RSD+CC+Pantheon$^+$ datasets for both $f(Q)$ and $\Lambda$CDM models using $\Gamma_2$ parametrization. The black dashed line represents the theoretical value of $\gamma_0=6/11$ for the $\Lambda$CDM model.}
\label{cont3}
\end{figure*}
\begin{figure*}[t]
    \centering
    \includegraphics[scale=0.32]{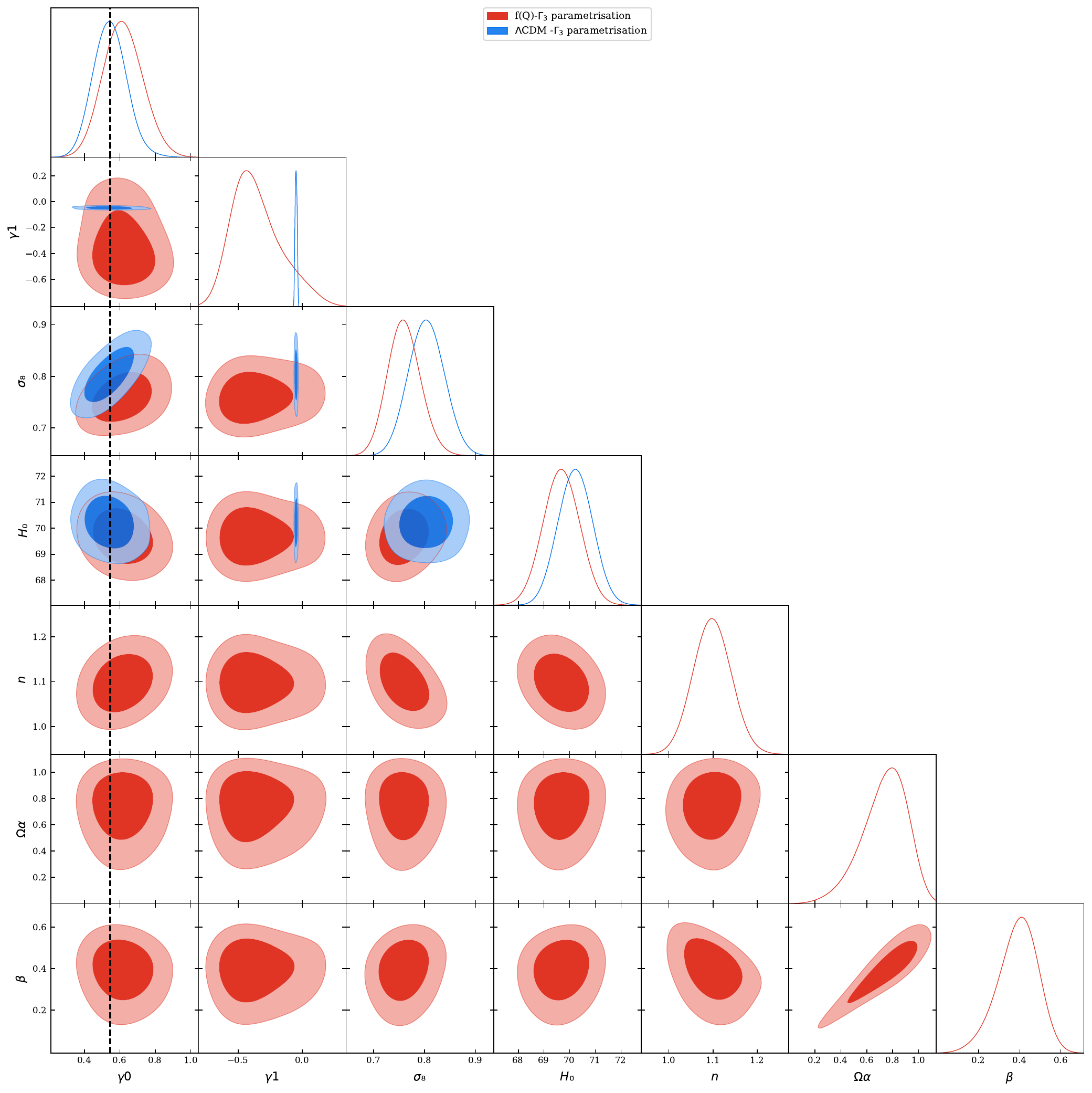}
    \caption{This figure shows the $1\sigma$ and $2\sigma$ confidence contours and the posterior distributions obtained from the RSD+CC+Pantheon$^+$ datasets for both $f(Q)$ and $\Lambda$CDM models using $\Gamma_3$ parametrization. The black dashed line represents the theoretical value of $\gamma_0=6/11$ for the $\Lambda$CDM model.}
\label{cont4}
\end{figure*}

\subsection{Information criteria}
%In order to make a comparaison between deferent models, with different numbers of parameters $N_P$, $\chi_{2}$ function is inssifisant to compare between deferent parametrization, thats why we use the information critaria as AIC and BIC. \\

The secondary objective of this study is to determine the most suitable model according to observational data. The smallest $\chi^2$ value stands for the 
perfect fit i.e. the most preferred model by data. However, a higher number of parameters can artificially improve the fit, leading to a smaller $\chi^2$, rendering it unreliable for model comparison. To address this issue, we employ the Akaike Information Criterion (AIC), a statistical tool which depends on the number of parameters, denoted as $\mathcal{N}$, to assess accurately the goodness of the fit of the studied models \cite{akaike1974new,AIC}\\

\begin{equation}
AIC = \chi^2_{min}+2\mathcal{N}.
\end{equation}

Additionally, we introduce the corrected Akaike Information Criterion, denoted as $AIC_c$, which depends on both the number of data points, represented by $k$ (equal to 1778 in our case), and the number of parameters, denoted as $\mathcal{N}$, as described in \cite{AIC}.

\begin{equation}
AIC_c = \chi^2_{min}+2\mathcal{N}+\frac{2\mathcal{N}(\mathcal{N}+1)}{k-\mathcal{N}-1},
\end{equation}
similar to AIC, a model with a smaller value of $AIC_{c}$ indicates that the model is better supported by the observation data. Hence, we calculate 

\begin{equation}
\Delta AIC = AIC_{\text{model}}- AIC_{\text{min}}.
\end{equation}

The model selection rule of $\Delta$AIC, is as follows: the models with $0< |\Delta AIC |<2$ have substantial support, those with $4< | \Delta AIC  |<7$ have considerably less support, and models with $ | \Delta AIC |>10$ have essentially no support, with respect to the reference model.  We also calculate the Bayesian Information Criterion (BIC), given by \cite{BIC}

\begin{equation}
BIC = \chi^2_{min}+\mathcal{N}\log{(k)},
\end{equation}
where a smaller value of BIC indicates that the model is better supported by the observation data. Therefore, we calculate
\begin{equation}
\Delta BIC =  BIC_{\text{model}}- BIC_{\text{min}}.
\end{equation}
The strength of evidence against the model with the highest BIC value is as follows: for  0$\leqslant\Delta$BIC$<$2   means that the evidence is insufficient, for 2$\leqslant\Delta$BIC$<$6, indicates  that there is a positive evidence, for 6$\leqslant\Delta$BIC$<$10 indicates a solid evidence.

\section{Resullts and duscussions} \label{Resullts}

%This section delves into the findings from our analysis of two expansion models: $\Lambda$CDM and $f(Q)$ gravity. We impose constraints on each model across four parameterizations $\Gamma_{0-3}$, using a combination of datasets: Pantheon$^{+}$ + CC + RSD.  Table \ref{Bf} presents the mean values and their associated errors at $1 \sigma(68 \%$ confidence level, CL) for the studied models. For the $\Lambda \mathrm{CDM}$ model, the free parameter vector $P_{\Lambda}$ consists of $\left(\Omega_m, \sigma_8, H_0, M, P_{\gamma} \right)$. while In $f(Q)$, the free parameter vector $P_{f(Q)}$ includes $\left(P_{Q},\sigma_8, H_0, M, P_{\gamma}\right)$. It is worth noting that for $\Gamma_{0}$, the associate growth index vector is $P_{\gamma}={\gamma_{0},0}$. while for $\Gamma_{1-3}$ the vector becames $P_{\gamma}={\gamma_{0},\gamma_{1}}$. In addition, Due to diferent numbre of free parameter between $\Lambda$CDM and $f(Q)$, we use AIC and BIC, which present in Table \ref{AICBIC}.  \\

In this section, we present the results obtained for $\Lambda$CDM and $f(Q)$ gravity models. We constraint the growth index parameters $(\gamma_{0},\gamma_{1})$ of each model using four parameterizations, $\Gamma_{0-3}$ described previously in the section \ref{matter perturbations}, Eqs (\ref{para0-3}), using the following combination of datasets: Pantheon$^{+}$, CC , and RSD. Table \ref{bf} summarizes the mean values and corresponding errors at the $1\sigma$ (68$\%$ confidence level, C.L.), while Figs. \ref{cont1}, \ref{cont2}, \ref{cont3}, and \ref{cont4} depict the confidence contour plots in two dimensions (2D) for the model parameters corresponding to $\Gamma_{0-3}$ for both models $\Lambda$CDM (in blue) and $f(Q)$ gravity (in red). The black dashed lines represent the theoretical value of $\gamma_{\Lambda}=\frac{6}{11}$ for $\Lambda$CDM\\

In the $\Lambda$CDM model, the parameter vector encompasses $\theta_{\Lambda}=\left(\Omega_m, \sigma_8, H_0, M, P_{\gamma} \right)$, whereas for $f(Q)$ gravity, the parameter vector comprises $\theta_{Q}=\left(P_{Q},\sigma_8, H_0, M, P_{\gamma}\right)$. Notably, for $\Gamma_{0}$, the associated growth index vector is represented as $P_{\gamma}\equiv \gamma_{0}$, while for $\Gamma_{1-3}$, it becomes $P_{\gamma}=(\gamma_{0},\gamma_{1})$. Additionally, considering the different number of free parameters between the models mentioned above, we employ the Akaike Information Criterion (AIC) and Bayesian Information Criterion (BIC), as elaborated in Table \ref{AICBIC}.

\subsection{Constant growth index}
\subsubsection{$\Gamma_0$ parametrization model}

Firstly, we consider the $\Gamma_0$ parametrization $\left(\gamma=\gamma_0\right.$ and $\gamma_1=0$); see Eq. (\ref{para0-3}), which implies that the corresponding statistical vector for the growth index becomes: $P_{\gamma} \equiv \gamma_{0}$.\\
Regarding the $\Lambda$CDM model using the $\Gamma_0$ parametrization, we find that the growth index $\gamma_{0}=0.545_{-0.096}^{+0.096}$ with $\chi_{min}^{2}=1604.15$ for $\mathcal{N}_{\Lambda}=5$ degrees of freedom. This result is in good agreement within $1\sigma$ errors with those of \cite{Specogna}, where they found $\gamma_{0}=0.536_{-0.018}^{+0.040}$ and $\gamma_{0} =0.558_{-0.018}^{+0.024}$ using \textit{ACT\footnote{Atacama Cosmology Telescope}}+WMAP+BAO and \textit{SPT\footnote{South Pole Telescope}}+WMAP+BAO, respectively. It is also consistent with \cite{Clustering of LRGs} within $1\sigma$ errors, which found $\gamma_{0}= 0.56_{-0.05}^{+0.05}$ using Clustering of \textit{LRGs\footnote{Luminous Red Galaxies}}+growth data, and with \cite{fr,ft}, which found $\gamma_{0}= 0.597_{-0.046}^{+0.046}$ using SnIa+BAO+WMAP+CMB+growth dataset. Therefore, the obtained value is in very good agreement with the theoretical value of $\gamma_{\Lambda}= 6/11$. It becomes evident that we do not restrict any parameter in our present analysis. By employing the most recent growth dataset together with Pantheon$^{+}$ and CC, we can place strong constraints on $\Omega_m = 0.273_{-0.011}^{+0.011}$, $\sigma_{8}=0.808_{-0.0359}^{+0.0359}$, and $H_{0}=70.228_{-0.661}^{+0.661}$, which are also in good agreement with \cite{Mhamdi2}.\\

 %--
%Concerning the $f(Q)$ expansion model, we find that $\gamma_{0} =0.571_{-0.110}^{+0.095}$  with $\chi_{min}=1599.61$  for $\mathcal{N}_{\Lambda}=7$ degrees of freedom. The obtained value of growth index consider somewhat greater from the $\gamma_{\Lambda}$ with differences of $4.68\%$. Additionally, the obtained cosmological value $P_{Q}=(0.72_{-0.12}^{+0.22}, 0.387_{-0.089}^{+0.11}, 1.093_{-0.042}^{+0.042})$, $H_{0}=69.708_{-0.697}{+0.697}$, and $\sigma_{8}=0.762_{-0.037}{+0.032}$ are in good agrement with \cite{Mhamdi2}.\\
Regarding the $f(Q)$ expansion model, we find that $\gamma_{0} =0.571_{-0.110}^{+0.095}$ with $\chi_{min}^{2}=1599.61$ for $\mathcal{N}_{Q}=7$ degrees of freedom. The obtained value of the growth index is somewhat greater than $\gamma_{\Lambda}$, with differences of $4.66\%$. Additionally, the cosmological values obtained, $P_{Q}=(0.72_{-0.12}^{+0.22}, 0.387_{-0.089}^{+0.11}, 1.093_{-0.042}^{+0.042})$, $H_{0}=69.708_{-0.697}^{+0.697}$, and $\sigma_{8}=0.762_{-0.037}^{+0.032}$, are in good agreement with \cite{Mhamdi2}. A comparison with $f(R)$ and $f(T)$ theories reveals that the growth index of the $f(Q)$ model is lower than that of the Starobinsky $f(R)$ model \cite{fr,AStarobinsky} and the power law $f(T)$ model \cite{ft,RBengochea}, with differences of $4\%$ and $5.28\%$, respectively.\\

%-- 
Due to the difference in the number of degrees of freedom between $\Lambda$CDM and $f(Q)$ ($\mathcal{N}_{\Lambda}\neq \mathcal{N}_{Q}$), we compute $AIC_{c}$ and $BIC$ for both models. The value of $AIC_{c,f(Q)}= 1613.67$ is smaller than the corresponding $\Lambda$CDM value $AIC_{c,\Lambda}= 1614.18$, indicating that the $f(Q)$ model, now appears to fit the RSD data slightly better than the usual $\Lambda$CDM. However, the small $|\Delta AIC|$ value (i.e., $\sim$ 0.51) points out that  $\Gamma_0$ parametrization within $f(Q)$ model has substantial support with respect to $\Lambda$CDM model. As for $\Delta BIC$ (i.e., $\sim$ 10 ), it reports less evidence against $\Gamma_0$ parametrization within $f(Q)$ gravity model.\\
%-- 

\begin{table*}[t!]
    \centering
    \begin{tabular}{|c|c|cccccc|}
        \hline
        Model &  Parametrization  & $\chi_{min}^{2}$ & $\chi_{red}^{2}$  &  $AIC_{c}$  &   $\left | \Delta AIC \right |$ &$ BIC$ & $\Delta$BIC \\
        \hline
                                 & $\Gamma_0$  &1604.15& 0.90476&1614.18&        0       & 1641.57   &   0   \\ 
                                 & $\Gamma_1$  &1604.07&0.90523&1616.12&        1.94         & 1648.97 &   7.399  \\
     $\Lambda$CDM      & $\Gamma_2$  & 1604.13& 0.90526 &  1616.18 &    2       &1649.03& 7.459 \\
                                 & $\Gamma_3$ &1604.09&0.90524& 1616.14 &        1.96        &  1641.57  &  7.419  \\    

        \hline
                               & $\Gamma_0$             &1599.61&0.90322&1613.67&     0.51     &    1651.99 &  10.423 \\
                              & $\Gamma_1$             &1599.55& 0.90370&1615.63&      1.45    &    1659.42  &   17.851    \\
                              $f(Q)$   & $\Gamma_2$   &  1599.49 & 0.90367&1615.57&   1.39   &  1659.36&      17.786    \\
                              & $\Gamma_3$             & 1599.48  &0.90366&1615.56&    1.38     &1659.35 &   17.779 \\
        \hline
    \end{tabular}
    \caption{This table shows values of $AIC$ and $BIC$ for $\Lambda$CDM and $f(Q)$ gravity models using $\Gamma_0$, $\Gamma_1$, $\Gamma_2$, and $\Gamma_3$ parametrizations.}
    \label{AICBIC} 
\end{table*}

\begin{figure}[h!]
    \centering
   \includegraphics[scale=0.6]{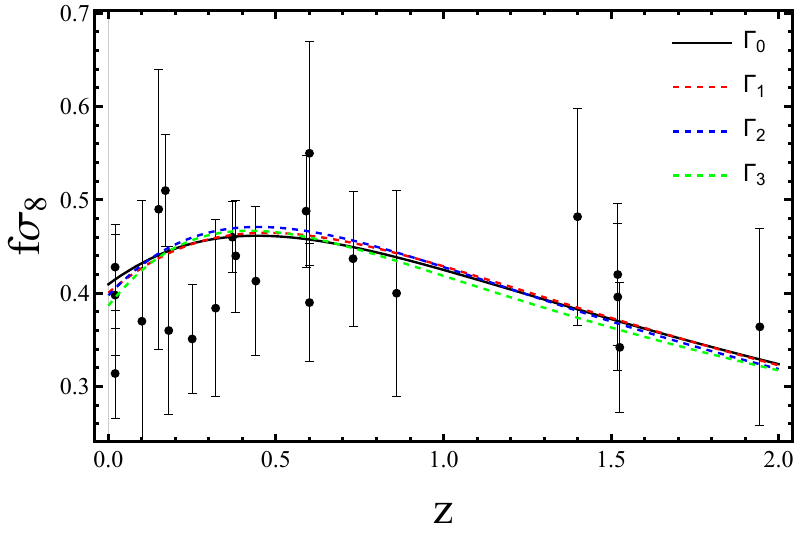}
    \caption{This figure shows the evolution of the predicted $f\sigma_8$ observable versus the redshift, $z$. The results given by the models $\Gamma_{0}$ (in black), $\Gamma_{1}$ (red dashed), $\Gamma_{2}$ (blue dashed), and $\Gamma_{3}$ (green dashed) in $f(Q)$ gravity.}
\label{fsigma8z}
\end{figure}

\begin{figure}[h!]
    \centering
    \includegraphics[scale=0.38]{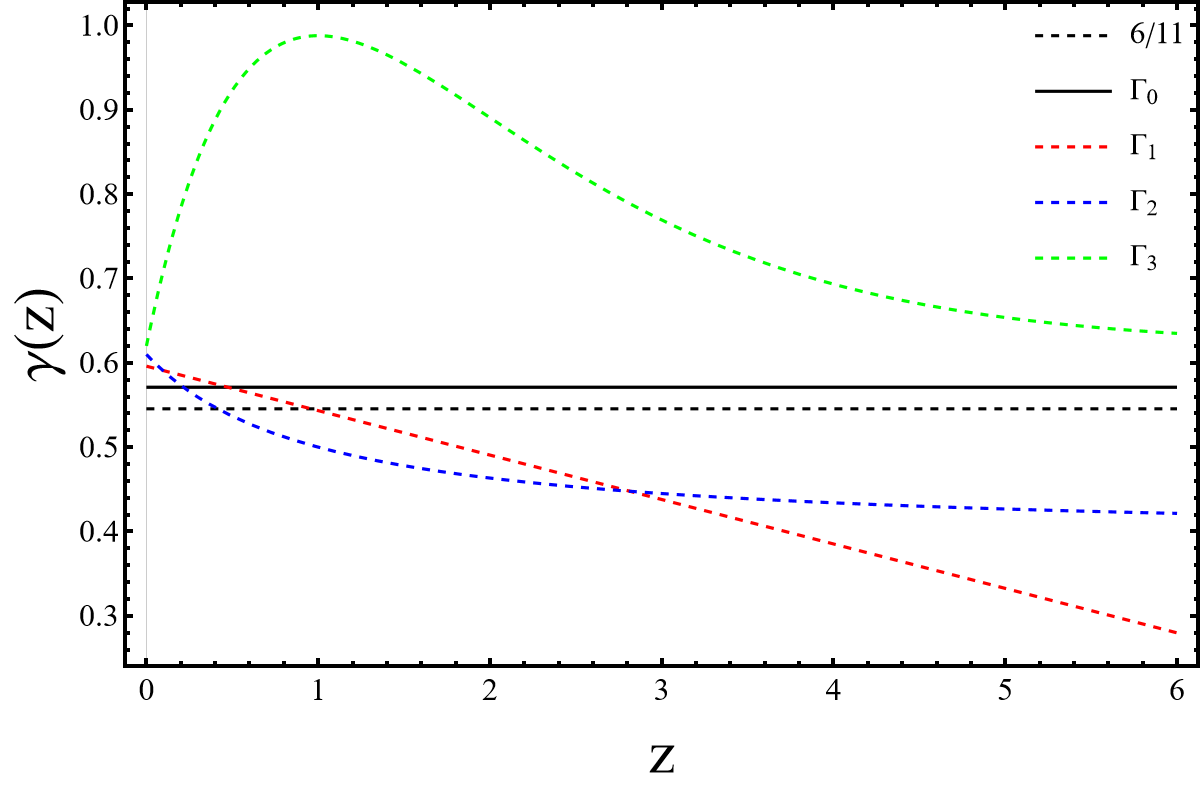}
    \caption{The evolution of the growth index function $\gamma$ versus redshift, $z$, for parametrization models $\Gamma_{0}$ (in black), $\Gamma_{1}$ (red dashed), $\Gamma_{2}$ (blue dashed), and $\Gamma_{3}$ (green dashed) in $f(Q)$ gravity. The black dashed line represents the theoretical value $\gamma_{\Lambda}=6/11$ of the $\Lambda$CDM model.}
\label{gammaz} 
\end{figure}

\subsection{Time varying growth index}
After presenting the simplest version of the growth index, $\Gamma_0$, it seems appropriate to discuss the observational constraints on the time-varying growth index, $\gamma(z)$, for the $\Gamma_{1-3}$ parametrization models. This discussion follows the considerations outlined in section \ref{matter perturbations}. Now, the corresponding statistical vector for the growth index becomes $P_{\gamma}=(\gamma_0, \gamma_1)$.

\subsubsection{$\Gamma_1$ parametrization model}

Concerning the $\Lambda$CDM model using the $\Gamma_1$ parametrization, we find that the growth index parameter results are $\gamma_{0}=0.547_{-0.096}^{+0.096}$ and $\gamma_{1}=-0.0106 _{-0.0014}^{+0.0014}$, with $\chi_{min}^{2}=1604.07$ for $\mathcal{N}_{\Lambda}=6$ degrees of freedom. This result is in good agreement within $1\sigma$ errors with those of \cite{fr,ft}, where they found $\gamma_{0}=0.567_{-0.066}^{+0.066}$ and $\gamma_{1} =0.116_{-0.191}^{+0.191}$ using SnIa+BAO+WMAP+CMB+growth dataset.\\
Concerning the $f(Q)$ model, we find that $\gamma_{0} =0.596_{-0.11}^{+0.097}$ and $\gamma_{1} =-0.0527_{-0.0098}^{+0.0057}$, with $\chi_{min}^{2}=1599.55$ for $\mathcal{N}_{Q}=8$ degrees of freedom. The obtained $\gamma_{0}$ value is greater than $\gamma_{\Lambda}$ by a difference of $9\%$. A comparison between $f(R)$ and $f(T)$ theories indicates that the growth index of the $f(Q)$ model is greater than that of both the Starobinsky $f(R)$ model \cite{fr,AStarobinsky} and the power law $f(T)$ model \cite{ft,RBengochea}, with differences of $5\%$ and $6.58\%$, respectively.\\

We notice that the value of $AIC_{c,f(Q)}= 1615.63$ is smaller than the corresponding $\Lambda$CDM value $AIC_{c, \Lambda}= 1616.12$, indicating that the $f(Q)$ model appears to fit slightly better than the usual $\Lambda$CDM. However, the small $|\Delta AIC|$ value (i.e., $\sim$ 1.45) indicates that $\Gamma_1$ parametrization within the studied $f(Q)$ model has 
substantial support with respect to $\Lambda$CDM. In the other hand, the value of $\Delta BIC=17.851$ reports strong evidence against $\Gamma_1$ parametrization within $f(Q)$ gravity model.

\subsubsection{$\Gamma_2$ parametrization model}

Regarding the $\Lambda$CDM model using the $\Gamma_2$ parametrization, we find that the growth index parameters are $\gamma_{0}=0.544_{-0.097}^{+0.097}$ and $\gamma_{1}=-0.00827_{-0.00154}^{+0.00154}$, with $\chi_{\text{min}}^{2}=1604.13$ for $\mathcal{N}_{\Lambda}=6$ degrees of freedom. This result is in good agreement within $1\sigma$ errors with those of \cite{fr,ft}, where the authors found $\gamma_{0}=0.561_{-0.068}^{+0.068}$ and $\gamma_{1} =0.183_{-0.269}^{+0.269}$ using SnIa+BAO+WMAP+CMB+growth dataset. Additionally, it is within $1\sigma$ errors of the results of \cite{Mehrabi}, where the authors found $\left(\gamma_0, \gamma_1\right) \simeq(0.557,-0.012)$ using SnIa+BAO+CMB+Big Bang Nucleosynthesis+$H(z)$+growth index dataset.\\

While in the $f(Q)$ model, we find that $\gamma_{0}=0.544_{-0.097}^{+0.097}$ and $\gamma_{1} =-0.0527_{-0.0098}^{+0.0057}$, with $\chi_{min}^{2}=1599.49$ for $\mathcal{N}_{Q}=8$ degrees of freedom. The obtained $\gamma_{0}$ value is greater than $\gamma_{\Lambda}$ with differences of $9\%$. When comparing the growth index of $f(R)$ and $f(T)$ theories with that of $f(Q)$, it becomes evident that the growth index of the $f(Q)$ model surpasses both the Starobinsky $f(R)$ \cite{fr,AStarobinsky} and the power law $f(T)$ models \cite{ft,RBengochea}, exhibiting differences of $8.36\%$.\\

We notice that the value of $AIC_{c, f(Q)}= 1615.57$ is smaller than the corresponding $\Lambda$CDM value $AIC_{c, \Lambda}= 1616.18$, indicating that the $f(Q)$ model appears to fit slightly better than the usual $\Lambda$CDM. However, the small $|\Delta AIC|$ value (i.e., $\sim$ 1.39) indicates that the $\Gamma_1$ parametrization within the studied $f(Q)$ model has substantial support with respect to $\Lambda$CDM. In the other hand, the value of $\Delta BIC=17.886$ reports a strong evidence against $\Gamma_2$ parametrization within $f(Q)$ gravity model.

\subsubsection{$\Gamma_3$ parametrization model }

In the context of the $\Gamma_3$ parametrization, concerning the $\Lambda$CDM model, we find that the growth index parameters are $\gamma_{0}=0.542_{-0.092}^{+0.092}$ and $\gamma_{1}=-0.047_{-0.007}^{+0.007}$, with $\chi_{min}^{2}=1604.09$ for $\mathcal{N}_{\Lambda}=6$ degrees of freedom. Meanwhile, in the $f(Q)$ model, we find that $\gamma_{0}=0.62_{-0.12}^{+0.10}$ and $\gamma_{1} =-0.359_{-0.24}^{+0.093}$, with $\chi_{min}^{2}=1615.56$ for $\mathcal{N}_{Q}=8$ degrees of freedom. The obtained $\gamma_{0}$ value is greater than $\gamma_{\Lambda}$ by approximately $9\%$.\\

As observed in previous $\Gamma_{0}$, $\Gamma_{1}$, and $\Gamma_{2}$ models, the value of $AIC_{c,f(Q)}= 1599.48$ is also smaller than the corresponding $\Lambda$CDM value $AIC_{c,\Lambda}= 1604.09$, indicating that the $f(Q)$ model appears to fit slightly better than the usual $\Lambda$CDM. However, the small $|\Delta$AIC$|$ value (i.e., $\sim$ 1.38) indicates that the $\Gamma_3$ parametrization within the studied $f(Q)$ model has substantial support with respect to $\Lambda$CDM. In the other hand, the value of $\Delta BIC=17.78$ points out to a strong evidence against $\Gamma_3$ parametrization within $f(Q)$ gravity model.\\

Let's delve into the cosmic evolution of the growth factor, $f\sigma_8(z)$, and the growth index, $\gamma(z)$, across various parametrization models, $\Gamma_{0-3}$, within the context of $f(Q)$ gravity. Our aim is to analyze and quantify their evolution as defined by Eqs. (\ref{f}) and (\ref{para0-3}). Fig. \ref{fsigma8z} illustrates the predicted evolution of $f\sigma_8$ using the mean values of $P_{Q}$ and $P_{\gamma}$ obtained previously from RSD data combined with Pantheon$^{+}$ and CC. (Table \ref{bf}). Notably, the predictions clearly align with the RSD measurements, indicating the effectiveness of each parametrization model. Therefore, Fig. \ref{gammaz} depicts the evolution of $\gamma(z)$, where the dashed black line represents the prediction of the $\Lambda$CDM model. Remarkably, for $\Gamma_{0}$, the growth index remains constant throughout the entire range $z\sim 0-6$. However, $\Gamma_{1}$ and $\Gamma_{2}$ exhibit medium deviations from the growth index $\gamma_{\Lambda}$, whereas $\Gamma_{3}$ displays significant deviations within the range of $z\sim 0.5-3$.

\section{Conclusion}\label{remarks}

It is widely acknowledged that the growth index, $\gamma$, plays a key role in understanding the formation and evolution of large-scale structures over time. Moreover, it serves as a valuable tool for testing theories of gravity, such as Einstein’s general relativity. In this work, we have used recent observational data, including Pantheon$^{+}$, cosmic chronometer, and redshift space distortion datasets, to constrain the parameters of the growth index within the framework of $f(Q)$ gravity. Our aim was twofold: first, apply a Markov Chain Monte Carlo analysis across various $\gamma$ parametrization models within the context of $f(Q)$ gravity, and second, to quantify the deviation between our $f(Q)$ and the standard $\Lambda$CDM models.\\

Initially, we have explored a constant growth index parametrization, $f(z)=\Omega_{m}(z)^{\gamma}$, where $\gamma=\gamma_{0}$. For $\Lambda$CDM, our analysis revealed that the observed growth index $\gamma_{0}=0.545\pm0.096$ aligns well, within $1\sigma$ errors, with results from various previous studies, such as \cite{Specogna, Clustering of LRGs, fr,ft}. Moreover, it is in very good agreement with the theoretically predicted value of $\gamma_{\Lambda}=11/6$. However, for the $f(Q)$ model, we have found a deviation of $4.66\%$ from the $\Lambda$CDM model, with the value of the growth index $\gamma_{0} = 0.571_{-0.110}^{+0.095}$.\\

Subsequently, we have considered a time-varying growth index parametrization, $f(z)=\Omega_{m}(z)^{\gamma(z)}$, with $\gamma(z)=\gamma_0+\gamma_1 y(z)$, where $y(z)=z, 1-1/(1+z),$ and $z e^{-z}$. We have found that $(\gamma_0,\gamma_1)=(0.596,-0.0527)$, $(\gamma_0,\gamma_1)=(0.61,-0.22)$, and $(\gamma_0,\gamma_1)=(0.62,-0.359)$, for $\Gamma_1$, $\Gamma_2$, and $\Gamma_3$ models, respectively. However, we have found that the $\gamma_{0}$ values for the $\Gamma_{1-3}$ parametrization models are greater than that of $\Lambda$CDM by a difference of 9\%.\\

Finally, an examination of the evolution of the growth index $\gamma(z)$ versus redshift $z$ for each parametrization model revealed distinct behaviors. While $\Gamma_{0}$ maintained constancy across the entire range $z\sim 0-6$, $\Gamma_{1}$ and $\Gamma_{2}$ exhibit medium deviations from the growth index $\gamma_{\Lambda}$, whereas $\Gamma_{3}$ displays significant deviations within the range of $z\sim 0.5-3$.

%Finally, we have the evolution of the growth index  

%\section{Appendix}

%\begin{table*}[t!]
  %  \centering
  %  \label{tab:comparison}
    %\begin{tabular}{c|c|c|c|c|c}
       % \toprule
       % Model & $\Gamma_{0}$  & $\Gamma_{1}$ &  $\Gamma_{2}$ & $\Gamma_{3}$ & References \\
         %\hline
        %$f(R)$ & $\gamma_{0}= 0.598$ & $(\gamma_0,\gamma_1)=(0.573,0.097)$ & $(\gamma_0,\gamma_1)=(0.579, 0.101)$ & -& \cite{fr} \\
           %      & $\gamma_{0}= 0.594$ & $(\gamma_0,\gamma_1)=(0.567, 0.113)$ & $(\gamma_0,\gamma_1)=(0.561, 0.179)$ & - &\cite{fr}\\
         %\hline
       %$f(T)$ & $\gamma_{0}=0.602$ & $(\gamma_0,\gamma_1)=(0.558, 0.187) $ & $(\gamma_0,\gamma_1)=( 0.564, 0.213)$&- & \cite{ft}\\
           % & $\gamma_{0}=0.596$ & $(\gamma_0,\gamma_1)=(0.566, 0.116) $ & $(\gamma_0,\gamma_1)=(0.561, 0.183)$&-& \cite{ft} \\
         %\hline
     %$f(Q)$   & $\gamma_{0}=0.571 $ & $(\gamma_0,\gamma_1)=(0.596,-0.053)$  & $(\gamma_0,\gamma_1)=(0.61,-0.220)$ & $(\gamma_0,\gamma_1)=(0.62,-0.359)$ & [Our study]\\
       % \bottomrule
   % \end{tabular}
 %\caption{Comparison between growth index of $f(R)$, $f(T)$, and $f(Q)$.}
%\end{table*}

\end{document}